\documentclass[a4paper,11pt]{article}
\usepackage{pos}
\usepackage[T1]{fontenc}
\usepackage[utf8]{inputenc}
\usepackage{caption}
\usepackage{subcaption}
\usepackage[nottoc]{tocbibind}

\usepackage{mathtools}
\usepackage{caption}
\newcommand{\oPs}{\mathrm{o-Ps}}
\newcommand{\pPs}{\mathrm{p-Ps}}
\DeclarePairedDelimiter\inprod\langle\rangle
\usepackage{commath}
\usepackage[numbers,sort&compress]{natbib}
\title{Physics Beyond the Standard Model with the J-PET detector}
\author*[a]{Elena P\'erez del R\'io}
\author{for the J-PET Collaboration}

\affiliation[a]{Faculty of Physics, Astronomy and Applied Computer Science, Jagiellonian University,\\
  S. Łojasiewicza 11, 30-348, Kraków, Poland}

\emailAdd{elena.rio@uj.edu.pl}
\abstract{The Positronium (Ps) system, a bound state of an electron and a positron, is suitable for testing the predictions of quantum electrodynamics (QED), since its properties can be perturbatively calculated to high accuracy and, unlike the hydrogen system, is not affected by finite size or QCD effects at the current experimental precision level. This makes the Ps atom a good laboratory to test the Fundamental Symmetries of Physics and also search for new particles not included in the Standard Model (SM) of physics.

On one hand, time reversal (T) and CP symmetry violation have never been observed in pure leptonic systems like the Ps atom, and the current experimental limits for CP and CPT violation in the decays of Ps is currently set at the level of 10$^{-3}$, while for C violation are at the level of 10$^{-7}$. These limits are still six and two orders of magnitude lower than the expected precision of 10$^{-9}$ set by photon-photon interactions. Secondly, experiments searching for invisible decays of the Ps triplet state, the ortho-Positronium (o-Ps), which mainly decays to three photons, are being conducted, since they are sensitive to new physics scenarios, e.g. Mirror Matter (MM), a suitable Dark Matter candidate, proposed to restore parity (P) violation. By performing a high precision measurement of the o-Ps lifetime, the accuracy of the present QED calculations can be tested and a search for the invisible decays of the o-Ps can be performed.

These studies are conducted with the novel total-body Positron-Electron Tomograph (PET) scanner at the Jagiellonian University. The J-PET is a large and high precise medical imaging tool, based on the plastic scintillators. It is a high acceptance multi-purpose detector optimized for the detection of photons from positron-electron annihilation and can be used in a broad scope of interdisciplinary investigation, e.g. medical imaging, life-time measurements, quantum entanglement studies with o-Ps, and tests of discrete symmetries.}

\FullConference{%
  *** The European Physical Society Conference on High Energy Physics (EPS-HEP2021), ***\\
  *** 26-30 July 2021 ***\\
  *** Online conference, jointly organized by Universität Hamburg and the research center DESY ***
}

\begin{document}
\maketitle

\section{Introduction}
%The positronium, a bound state of electron and positron  is a unique system to perform high precise tests, due to its simplicity. Being a system of lepton and antilepton, its properties are precisely described by Quantum Electrodynamics (QED) with very small radiative corrections from Quantum Chromodynamics (QCD) and weak interaction effects in the SM. The ground state of the Ps has two possible configurations, a singlet state $^1S_0$ para-positronium, p-Ps, where the spins of the electron and the positron are antiparallel, and the $^3S_1$ triplet, orhto-positronium, where the spins are parallel. The lifetimes of the two ground states are 125 picosecond for the p-Ps and 142 nanoseconds in the case of o-Ps. Reviews of Ps physics can be found in ~\cite{cassidy, karshenboim, gninenko}. Due to its characteristics, positronium is an eigenstate of the $\mathcal{P}$ operator, and $\mathcal{C}$ as well.

The positronium (Ps), a bound state of electron and positron  is a unique system to perform high precise tests, due to its simplicity. Being a system of lepton and antilepton, its properties are precisely described by Quantum Electrodynamics (QED) with very small radiative corrections from Quantum Chromodynamics (QCD) and weak interaction effects in the SM. The ground state of the Ps has two possible configurations, a singlet state $^1S_0$ para-positronium, $\pPs$, and the $^3S_1$ triplet, orhto-positronium, $\oPs$. The lifetimes of the two ground states are 125 picosecond for the $\pPs$ and 142 nanoseconds in the case of $\oPs$. Due to its characteristics, Ps is an eigenstate of the $\mathcal{P}$ operator, as well as of $\mathcal{C}$.

The parity of Ps, in its ground state, is equal to $\lambda_{P} = -1$, while the charge conjugation is dependent on the spin S of the system, $\lambda_{C} = (-1)^{S}$. This implies that the $\pPs$ is a $\mathcal{CP}$-odd state while the $\oPs$ is even with respect to this operator. According to the Standard Model predictions, photon-photon interaction or weak interaction can mimic the symmetry violation at the level of $10^{-9}$ and $10^{-13}$, respectively~\cite{sozzi,Bernreuther,arbic, pokraka}. Thus, there is a range of about 7 orders of magnitude for possible observation of $\mathcal{CP}$ and/or $\mathcal{CPT}$ symmetry violation with respect to the existing limits~\cite{yamazaki,vetter2}. A proposed test of invariance under a certain operation is the measurement of non-vanishing expectation values of certain operators odd under a given transformation~\cite{gajos}. For the decay $\oPs \rightarrow 3\gamma$, they can be constructed from momenta of the final-state photons, their polarization and spin of the Ps~\cite{moskal}. Some of these operators are listed in~\ref{tab:table1}. The $\vec{k_1}$, $\vec{k_2}$, and $\vec{k_3}$ denote the momenta of the three photons in ascending order according to their moduli, and the vectors $\vec{\epsilon_1}$, $\vec{\epsilon_2}$, and $\vec{\epsilon_3}$ are their corresponding polarization vectors. For elemental positron sources the spin direction can be determined taking advantage of the fact that due to the parity violation of the beta decay, the positrons are longitudinally  polarized along their velocity direction. This polarization is preserved to high extend during the Ps formation~\cite{moskal,coleman}. Up to recently tests were only performed for the $\mathcal{CP}$ and $\mathcal{CPT}$ symmetries, yielding the mean values $\inprod{\vec{S}\cdot\vec{k_1}[\vec{S}\cdot(\vec{k_1}\times\vec{k_2})]} = 0.0013 \pm 0.0012$~\cite{yamazaki} and $\inprod{\vec{S}\cdot(\vec{k_1}\times\vec{k_2})}= 0.0071 \pm 0.0062$~\cite{vetter2}, respectively, and leaving six orders of magnitude of unexplored precision.

\begin{table}[!ht]
    \centering
    \caption{Discrete symmetry odd-operators using spin orientation of the o-Ps, polarization and momentum directions of the annihilation photons~\cite{moskal}.}
    \label{tab:table1}
    \begin{tabular}{|l|c|c|c|c|c|}
    \hline
         Operator & C & P & T & CP & CPT  \\
         \hline
         $\vec{S}\cdot\vec{k_1}$& + & - & + & - & - \\
         \hline         $\vec{S}\cdot(\vec{k_1}\times\vec{k_2})$& + & + & - & + & - \\
         \hline $(\vec{S}\cdot\vec{k_1})\cdot(\vec{S}\cdot(\vec{k_1}\times\vec{k_2}))$& + & - & - & - & + \\
         \hline
         $\vec{\epsilon_{1}}\cdot{\vec{k_{1}}}$ & + & - & - & - & +\\
         \hline
         $\vec{S}\cdot\vec{\epsilon_{1}}$ & + & + & - & + & - \\
         \hline
         $\vec{S}\cdot(\vec{k_{2}}\times\vec{\epsilon_{2}})$ & + & - & + & - & -\\
         \hline
    \end{tabular}
\end{table}
\subsection{The J-PET detector}

The J-PET detection system, J-PET (stands for The Jagiellonian-PET), is the first Positron Emission Tomography scanner built from plastic scintillators~\cite{jpet1, jpet2, jpet3}. %The characteristics of the plastic scintillators, together with the fully digital front-end electronics make the J-PET system a high performance detector with high timing resolution~\cite{sharma} that can be used in a broad spectrum of interdisciplinary investigations. 

\begin{figure}
\centering
\begin{minipage}{.5\textwidth}
  \centering
  \includegraphics[width=.6\linewidth]{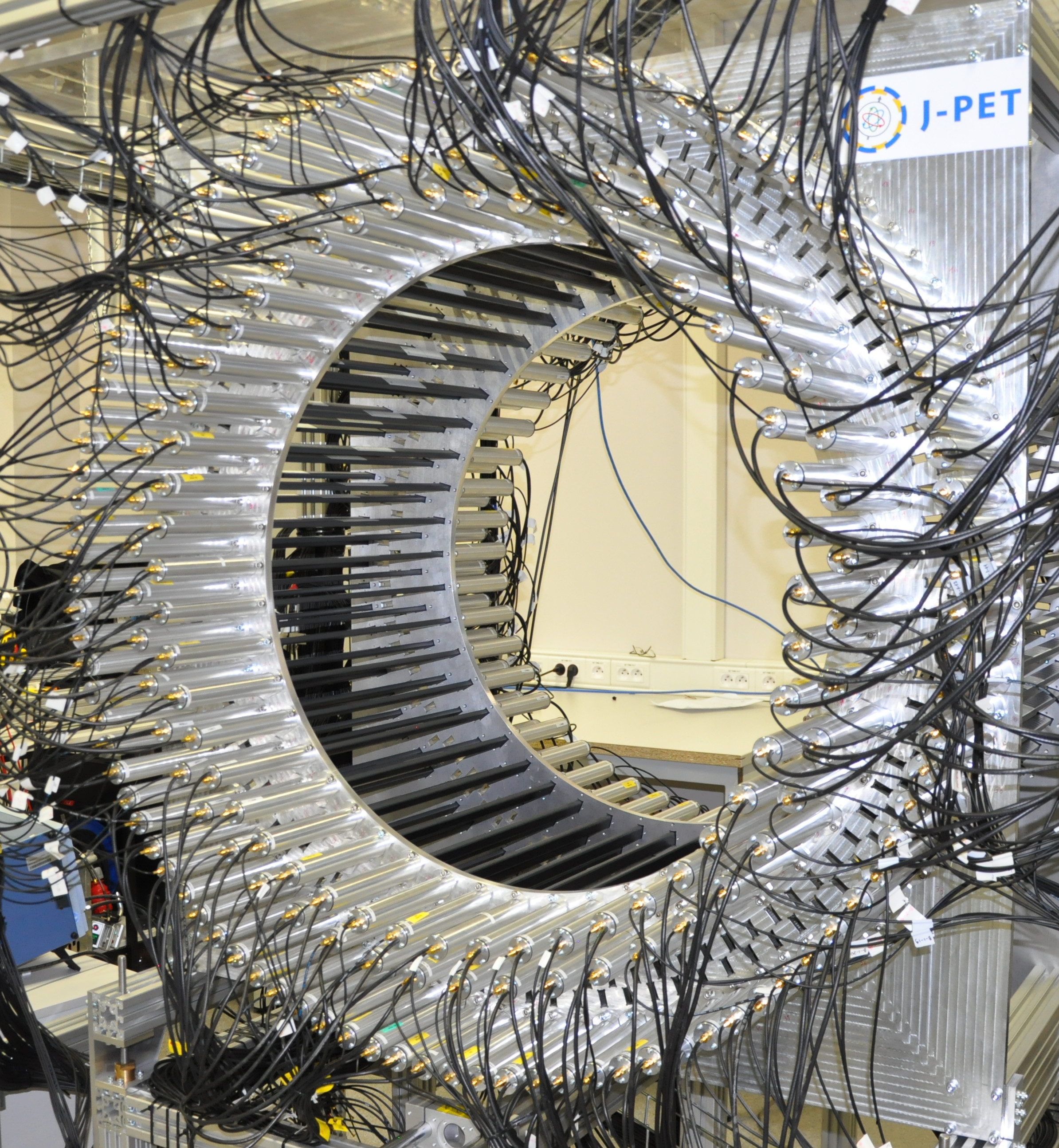}
  \captionof{figure}{The J-PET detector setup.}
  \label{fig:lab}
\end{minipage}%
\begin{minipage}{.5\textwidth}
  \centering
  \includegraphics[width=.6\linewidth]{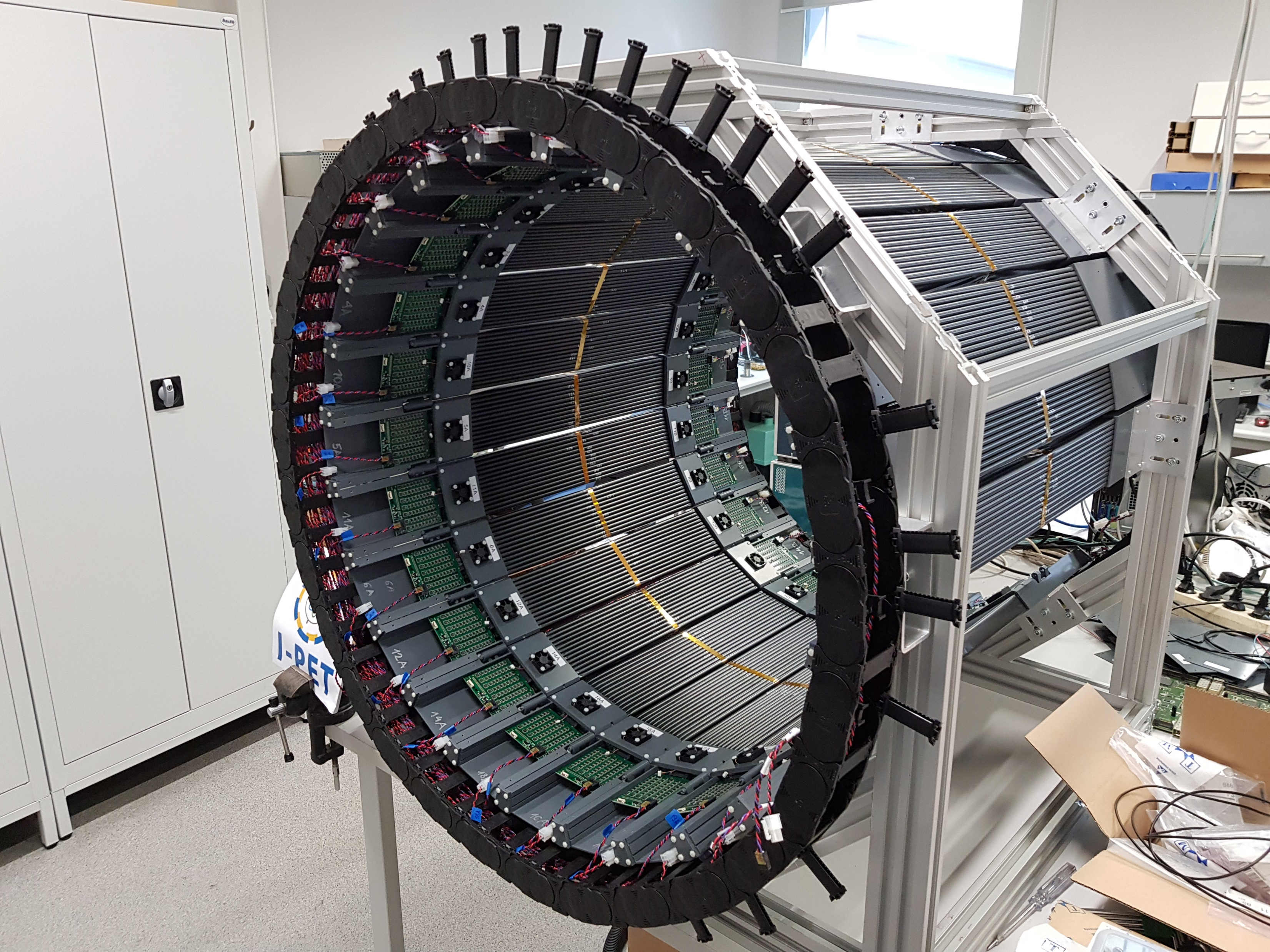}
  \captionof{figure}{The fourth layer of the J-PET detector. }
  \label{fig:layer4}
\end{minipage}
\end{figure}

J-PET at the present stage, see Fig.~\ref{fig:lab}, is built from three cylindrical layers including in total $192$ plastic scintillators strips with dimensions of $7\times 19 \times 500~\mathrm{mm}^3$. Light signals from each strip are converted to electrical signals by photomultipliers placed at opposite ends of the strip~\cite{NIM2015}. The position and time of the photons interacting in the detector material are determined based on the arrival time of light signals at both ends of the scintillator strips. The signals are probed in the voltage domain with the accuracy of about $30~\mathrm{ps}$ by a newly developed type of front-end electronics~\cite{Paka2017} and the data are collected by the trigger-less and re-configurable data acquisition system~\cite{Korcyl2016,Korcyl2018}.

%\begin{figure}[!ht]
%    \centering
%    \includegraphics[width=0.5\textwidth]{layer4_JPET.jpeg}
%    \caption{The fourth layer of the J-PET detector. It consists of $12$ plastic scintillator strips with dimensions $6\times 24 \times 500~\mbox{mm}^3$ read out by silicon photomultipliers. The inclusion of layer four will improve the time resolution by about a factor of $1.5$.}
%    \label{fig:layer4}
%\end{figure}
% WK -> minor thing, but you could remove older reference and add SoftwareX
For the data processing and simulations a dedicated software framework was developed~\cite{Krzemien2016,Krzemien2015,Krzemien2015framework}. The hit-position and hit-time are reconstructed by the dedicated reconstruction methods based on the compressing sensing theory and the library of synchronized model signals~\cite{Raczynski2014,Raczynski2015,Raczynski2017,NIM2015}. 

 New detection modules were designed and commissioned, in Fig.~\ref{fig:layer4}.  This fourth layer is read out by matrices of silicon photomultipliers (SiPM), which is expected to triple the efficiency for the single photon detection and improve the time resolution by about a factor of $1.5$~\cite{timing}.%A single module of the new detection layer of the J-PET tomograph is an independent detection unit which sends information about timing of the signals via optical links to the data acquisition boards. It consists of 13 BC-404 plastic scintillator strips. Each strip has dimensions of 6x24x50 mm$^3$. %Scintillators are wrapped in Vikuiti ESR and Pokalon 100B foils. 
 %Each side of scintillator strip is read out by SiPMs (Silicon PhotoMultipliers) and analog signals are then fed into LVDS buffer of FPGA chip.
 
In order to produce Ps atoms, a point-like $^{22}Na$ source is placed in the center of the detector and is surrounded with XAD-4 porous polymer~\cite{xad-4}. The porous polymer enhances the $\oPs$ formation probability. Additionally, one of the unique features of the J-PET detector is its ability to measure the polarization of the annihilation photons, which can be determined as the vector product of the momenta of the photon before and after the scattering in the detector. 
\subsection{Test of CPT symmetry in ortho-positronium}
The J-PET collaboration has performed a search for a CPT-violating angular correlation in the three-photon
annihilation of $\oPs$ atoms. In a recent publication~\cite{nature2021}, we have measured the expectation value of the angular correlation $O_{CPT} = \frac{\vec{S}\cdot(\vec{k_1}\times\vec{k_2})}{\abs{\vec{k_1}\times\vec{k_2}}} = \cos{\theta}$, achieving a sensitivity three times better than the previous result~\cite{vetter2}. The newly image reconstruction technique described in the publication~\cite{nature2021}, and later in~\cite{imaging1, imaging2}, enables the use of the Ps polarization without the application of a magnetic field.

\begin{figure}
\centering
\begin{minipage}{.5\textwidth}
  \centering
  \includegraphics[width=.8\linewidth]{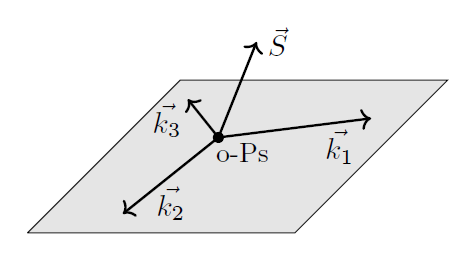}
  \captionof{figure}{Scheme of the decay plane from the $3\gamma$ annihilation and the spin of the Ps. Figure taken from~\cite{nature2021}.}
  \label{fig:plane}
\end{minipage}%
\begin{minipage}{.5\textwidth}
  \centering
  \includegraphics[width=.9\linewidth]{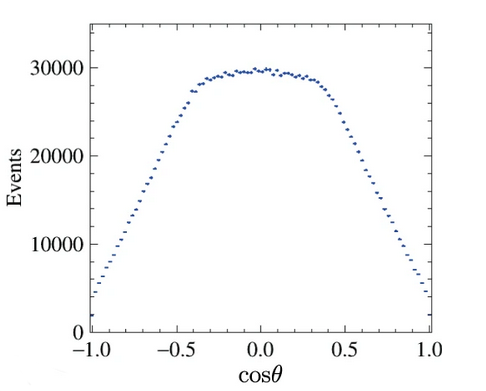}
  \captionof{figure}{Histogram of the determined $\cos{\theta}$ distribution. Figure from~\cite{nature2021}.}
  \label{fig:resultCPT}
\end{minipage}
\end{figure}

Using a $^{22}Na$ source of 10 MBq, the J-PET collected a sample of a total of $7.3 \times 10^6$ event candidates from the $\oPs \rightarrow 3\gamma$ annihilation process. For every selected event, the o-Ps spin axis is reconstructed, which allows to calculate the cosine of the angle formed between the spin, $\vec{S}$, and plane defined by the annihilation decay gammas, $\vec{k_1}\times\vec{k_2}$, the distribution of which is sensitive to CPT-violating asymmetries. The scheme of the measurement is presented in Fig.~\ref{fig:plane}.
The resulting distribution, shown in Fig.~\ref{fig:resultCPT}, for the $\cos{\theta}$, has been obtained with $1.9 \times 10^{6}$ identified $\oPs \rightarrow 3\gamma$ events.
The expectation value of the CPT-sensitive operator,$\inprod{O_{CPT}}$, needs to be corrected by the analysing power of the setup, which is dominated by the effective $\oPs$ polarisation, arising from the evaluation of the spin axis orientation in event-by-event basis. The total effective polarisation was estimated to be $P \sim 37.4\%$. We don't observe significant asymmetry over the entire region and obtain a value of:
\begin{equation}
    \inprod{O_{CPT} = 0.00025 \pm 0.00036}
\end{equation}
with the error dominated by the statistical uncertainty of 0.00033. Further measurements with improved detector setup and higher statistics will improve the sensitivity.

\subsection{Precision test in CP- and T-symmetry in the leptonic sector}
So far, no CP-violation has been observed in Ps decays, with the best limit having a sensitivity of $2.2\times10^{-3}$~\cite{yamazaki}. The present sensitivity was reached by measuring the angular correlation of $(\vec{S}\cdot\vec{k_1})\vec{S}\cdot\vec{k_1}\times\vec{k_2}$. In the J-PET experiment, the measurement without external magnetic field for Ps polarization can be simplified with a modified choice of the measured T-odd operator. A spin-independent operator is constructed for the $\oPs \rightarrow 3\gamma$ annihilations by including the polarization vector of one of the final state photons:
\begin{equation}
    C_{T}\prime = \vec{k_2}\cdot\vec{\epsilon_1}
\end{equation}

\begin{figure}[!ht]
    \centering
    \includegraphics[width=0.5\textwidth]{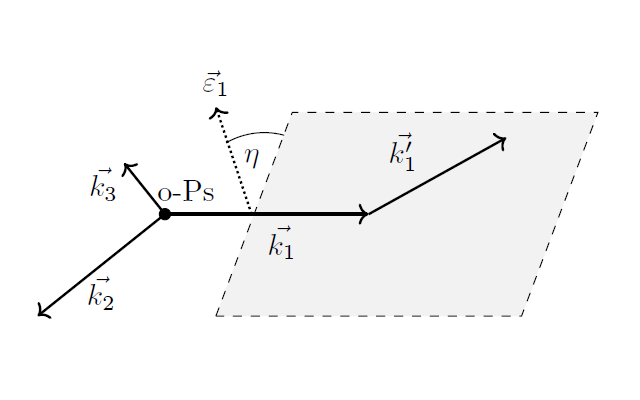}
    \caption{Scheme of estimation of polarization vector for a photon produced in $\oPs\rightarrow 3\gamma$ at J-PET. From~\cite{gajos}}
    \label{fig:toddMeas}
\end{figure}

where $\epsilon_1$ denotes the electric polarization vector of the most energetic photon and $\vec{k_2}$ is the momentum of the second most energetic one. Such angular correlations using photon electric polarization have been never explored in the $\oPs$ decays. In the J-PET, the ability of recording secondary interactions of scattered photons, as shown in Fig.~\ref{fig:toddMeas}, allows the measurement of $\inprod{C_{T}\prime}$. A preliminary result with a precision of $10^{-4}$ has been established with J-PET. Further data and the upgraded setup can increase the sensitivity at the level of $10^{-5}$.

\subsection{Mirror Matter searches in o-Ps}

The invisible decays of the o-Ps system are sensitive to new physics 
scenarios, e.g.\ mirror matter, 
milli-charged particles and extra space-time dimensions
\cite{Gninenko:2006sz}. Mirror matter was first proposed in connection with parity violation, suggesting that under spatial inversion, particles transform into a parity reflected new mirror particle state \cite{Lee:1956qn,Salam:1957,Okun:1966}, thus restoring parity symmetry in nature. Since this mirror partners interact mainly through gravity with the SM particles, they become dark matter candidates.
%The mirror particles would interact with normal particles mainly through gravity, becoming dark matter candidates. 

The o-PS decay rate within QED has been evaluated to two-loop level~\cite{Adkins:2002fg}. %
A detailed discussion can be found in e.g.~\cite{steven1}. 
The QED decay rate prediction yields $\Gamma = 7.039 979 (11) \times 10^6 s^{-1}$.\label{eq:qed_decay}
This decay rate is know with an accuracy of $10^{-6}$ including the logarithmic term, and $10^{-4}$ accounting only for the quadratic term, which roughly corresponds to the current best experimental accuracy of $ 2 \cdot 10^{-4}$~\cite{kataoka, vallery}. To reach the statistical uncertainty below  $10^{-4}$ one needs to reconstruct $10^{9}$ o-Ps events~\cite{crivelli:2018}. Taking into account the estimated number of o-Ps generated using the J-PET setup ($\approx 10^{13}$) and the efficiency for the detection of annihilation photons (2\%) and of the de-excitation photon (about 20\%), a ten times better sensitivity (below $10^{-5}$) can be reached after two years of data taking.  Presently, only the J-PET detector is able to perform such a sensitive measurement of the o-PS lifetime~\cite{kamil}. A more detailed description of the method, estimations and calculations can be found in~\cite{elena}.

\bibliography{biblio}

\end{document}